\documentclass[12pt,twoside]{article}
\usepackage{latexsym, amsthm, amsmath, bbm}
\usepackage{amssymb}
\usepackage{amsfonts}
\usepackage{amstext}
\usepackage{multicol}
\usepackage[table]{xcolor}
\usepackage{graphicx}
\usepackage{grffile}

\usepackage{subcaption}
\usepackage{hyperref}

\usepackage{authblk}

\theoremstyle{remark}

\usepackage{bm}
\usepackage{MnSymbol}

\extrafloats{100}

\title{States of Disarray: Cleaning Data for Gerrymandering Analysis}
\author[1]{Ananya Agarwal}
\author[1]{Fnu Alusi}
\author[1]{Arbie Hsu}
\author[1]{Arif Syraj}
\author[1]{Ellen Veomett}
\affil[1]{University of San Francisco, Department of Computer Science}

\begin{document}

\maketitle

\begin{abstract}
The mathematics of redistricting is an area of study that has exploded in recent years.  In particular, many different research groups and expert witnesses in court cases have used outlier analysis to argue that a proposed map is a gerrymander.  This outlier analysis relies on having an ensemble of potential redistricting maps against which the proposed map is compared.  Arguably the most widely-accepted method of creating such an ensemble is to use a Markov Chain Monte Carlo (MCMC) process.  This process requires that various pieces of data be gathered, cleaned, and coalesced into a single file that can be used as the seed of the MCMC process.

In this article, we describe how we have begun this cleaning process for each state, and made the resulting data available for the public \cite{eveomett-states}.  At the time of submission, we have data for 22 states available for researchers, students, and the general public to easily access and analyze.  We will continue the data cleaning process for each state, and we hope that the availability of these datasets will both further research in this area, and increase the public's interest in and understanding of modern techniques to detect gerrymandering.
\end{abstract}

\section{Democracy, Datasets, and Dirty Files}\label{sec:introduction}

In this polarized time, there is one topic that many across both sides of the aisle can agree on: we hate gerrymandering!  So what is this mutual enemy?  To understand the enemy, first we must understand the context of redistricting --- the drawing of a new districting map.  Redistricting occurs after the Census: the decennial enumerating of the US population, and where those residents live.  The reason redistricting happens after the Census is so that new districts can be drawn with nearly equal population; this is the idea of ``one person, one vote.''  Indeed, a districting map determines the regions of nearly equal population (districts) that vote for a politician (usually a single politician) to represent that region in a legislative body.  Having districts of nearly equal population ensures that each person's vote counts ``equally'' in determining their own representative.  We can now define our enemy:  gerrymandering is generally understood to be the result of drawing a map of districts in a way to disenfranchise a group of voters.  

Of course, exactly which maps disenfranchise which group of voters is not always (ever?) obvious to the naked eye, which is where mathematicians and computer scientists come in.  In recent years, there has been a great deal of excellent research into redistricting and gerrymandering by groups such as the Metric Geometry Gerrymandering Group (MGGG) \cite{mgggWebpage} and Duke's Quantifying Gerrymandering group \cite{quantifyingGerrymanderingWebpage}.  Additionally, computational experts have served as expert witnesses in gerrymandering cases \cite{EW_ClarkevWisconsin, SS_PSC_MT}, using data, statistics, and theory to argue the presence of gerrymandering in a map.  

This lively interest in addressing the problem of gerrymandering has not only captured computational researchers, but has grabbed the public's attention in many significant ways.  Some states have recently enacted new legislation regarding the redistricting process.  For example, Missouri passed a citizen's initiative Ammendment 1 in 2018 \cite{MO_first_ballot_measure}, and then another initiative Ammendment 3 in 2020 \cite{MO_second_ballot_measure}, both of which substantially changed the redistricting process.  And states such as Colorado \cite{CO_commission} and Michigan \cite{Mich_commish} have new independent redistricting commissions which drew redistricting maps after the 2020 Census.  In Ohio, citizens submitted their own redistricting maps \cite{OhioCompetition}, and at least one such citizen was frustrated that these maps were not taken more seriously \cite{OhioOpEd}.  

The public can directly engage with the redistricting process due, in large part, to the many tools that have been made available.  These tools include public-facing apps such as Dave's Redistricting App \cite{DRA} and Planscore \cite{PlanScore}, which allow the public to create and evaluate maps using simple metrics.  The public also has access to the powerful tools that researchers and expert witnesses typically use, such as the Python library gerrychain \cite{GerryDetails}.  The gerrychain library can create an ensemble of potential redistricting maps through a Markov Chain Monte Carlo (MCMC) process, and compare a proposed map to the other maps in the ensemble.  These kinds of tools are generally accepted as more reliable and informative than metrics \cite{GameabilityStudy}.

Our novel contribution to this area is producing a large number of datasets that can be used for these more advanced MCMC analyses, such as can be done with the Python library gerrychain.  The construction of these datasets was inspired by the MGGG's datasets \cite{MGGGstates}, which contain population, partisan, demographic, and districting data for the post-2010-Census districting maps.   The datasets we have constructed are the post-2020-Census versions of those datasets.  We note that this data coalescing and cleaning project could not have been done without the excellent work of the Redistricting Data Hub \cite{RDH}, Voting and Election Science Team (VEST) \cite{VEST} (whose data we used); and the MGGG's maup Python library contributors \cite{maup} (used to clean and coalesce the data).  We hope that, by making these datasets publicly available, we will further contribute to these  groups' work in supporting research and transparency.

In the remainder of Section \ref{sec:introduction}, we explain the MCMC process that is used in gerrychain and why the data needs to be cleaned as it has been.  We also detail the data we used from the Redistricting Data Hub's site (which includes VEST data).  In Section \ref{sec:repositories}, we describe why and how our datasets were constructed, and the details of what is available in the datasets.  Finally, we include some stories from the trenches, along with caveats and conclusions, in Section \ref{sec:caveats_conclusions}.

\subsection{The Structure Beneath}\label{sec:oa_and_gerrychain}

\begin{figure}[h]
\centering
\includegraphics[width=3in]{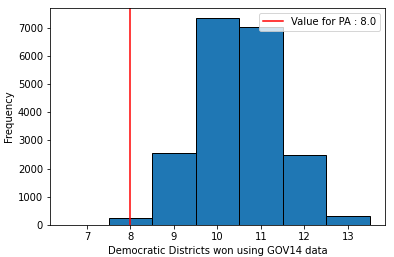}
\caption{Histogram of the number of Democratic districts won, using a gerrychain ReCom ensemble with 20,000 steps.  State is Pennsylvania, and election data is from the 2014 Gubernatorial election (election data used as a proxy for party preference).  The red bar corresponds to the number of districts the Democratic party would have won using the congressional map enacted in 2012.}
\label{fig:ensemble}
\end{figure}

Outlier analysis is one of the more widely accepted techniques of determining whether or not a proposed map is a gerrymander \cite{RamachandranGoldOutlierAnalysis}.  The idea behind outlier analysis is to construct an ensemble of potential redistricting maps, all of which satisfy whatever requirements a state may have for its own districts.  A proposed map is then compared against this ensemble; if this proposed map is an \emph{outlier}, this might signal that it was drawn with partisan intent.  For example, we used gerrychain's ReCom method to construct an ensemble of 20,000 potential congressional redistricting maps for Pennsylvania\footnote{These maps use the 2012-enacted map as the seed map, and thus have 18 districts each.}.  For each map, we used a single election (the 2014 Gubernatorial election) as a proxy for party preference, and counted how many districts the Democratic party would have won with each different map, and that same election data.  The results can be seen in Figure \ref{fig:ensemble}; the red bar corresponds to the number of districts the Democratic party would have won using the 2012-enacted map.  It seems low, compared to the vast majority of other maps in the ensemble, doesn't it?  This isn't a surprise, since the Pennsylvania Supreme Court tossed out that map as a Republican gerrymander in 2018 \cite{LeagueofWomenVoters2018}.

In order to implement an outlier analysis, one must construct the ensemble of comparison maps.  The MGGG \cite{mgggWebpage} and Quantifying Gerrymandering Group \cite{quantifyingGerrymanderingWebpage} (among others) use a Markov Chain Monte Carlo (MCMC) process to create these ensembles, and the MGGG has even created the Python library gerrychain \cite{gerrychain} to allow researchers, students, and the public to do this easily.

\begin{figure}[h]
\centering
\includegraphics[width=3in]{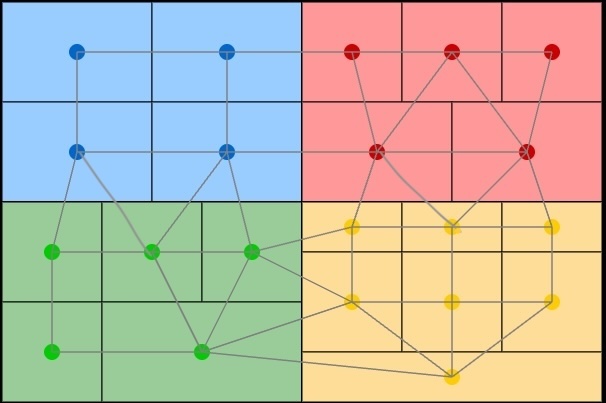}
\caption{Fake state with four districts (blue, red, green, and yellow), and blocks within each district.  Dual graph of all blocks is overlaid, nodes color-coded by district.}
\label{fig:state_districts}
\end{figure}

In order to motivate our data cleaning project, we briefly explain this MCMC process, in particular the ReCom method \cite{RecomMGGG} used by MGGG and implemented in gerrychain.  Since each district must have nearly equal population, and the smallest regions where Census data are tabulated are Census blocks, any districting map corresponds to census blocks which have been grouped together into districts\footnote{For the purposes of constructing the ensemble, we use election precincts instead of Census blocks.  We explain why in Section \ref{sec:data}, and state some issues with that choice in Section \ref{sec:caveats_conclusions}.}.   Thus, we consider the \emph{dual graph} of the Census blocks: each census block is a node, and two census blocks have an edge between them if they have a boundary of positive length.  In this context, a districting map is a coloring of nodes (each color a district) such that the subgraph induced by all nodes of the same color is connected (although there may be additional restrictions required by the state, besides just contiguity).  For a visualization of this process, see Figure \ref{fig:state_districts}.

\begin{figure}[h]
\centering
\includegraphics[width=3in]{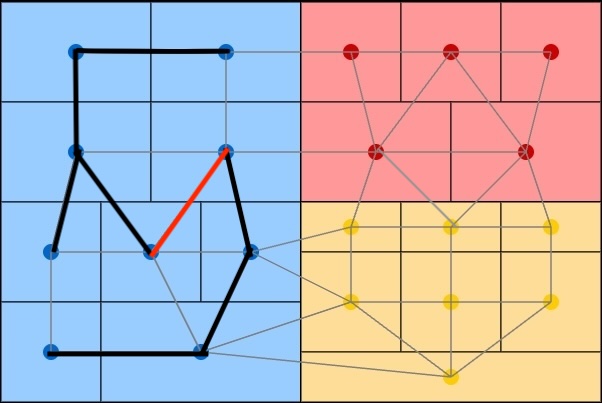}
\includegraphics[width=3in]{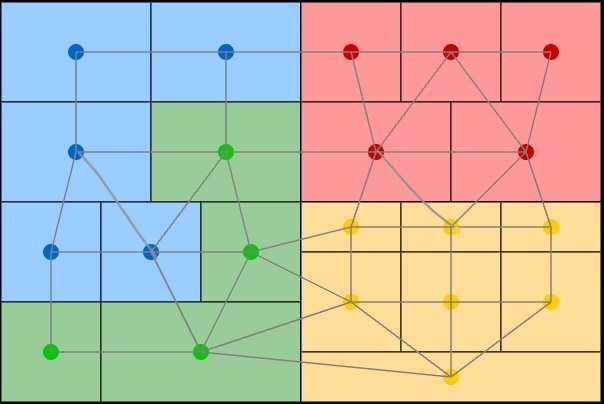}
\caption{Green and blue districts from Figure \ref{fig:state_districts} merged, and spanning tree constructed.  Black and red edges together are the spanning tree, with red edge being the chosen cut edge.  The resulting new districting map is on the right.}
\label{fig:merge_split}
\end{figure}

The ReCom MCMC process implemented in the python library gerrychain constructs an ensemble of maps as follows: first it finds two districts which are adjacent, and merges their dual graph into the dual graph of a doubly large district (doubly large in population).    The ReCom process then creates a random spanning tree of this graph, and checks each edge to see if cutting that edge results in two districts whose populations are within some threshold of the ideal district population.  We can see this in our fake state in the left image of Figure \ref{fig:merge_split}.   If there are several such cut edges, it chooses one of them uniformly at random to cut, which results in two new districts.  These two new districts, along with all of the other old districts,  is a new districting map!  The right side of Figure \ref{fig:merge_split} shows the resulting map for our fake state.  This process is repeated over and over in order to construct the ensemble.  

In order to create the dual graph used for the ReCom MCMC process, we need detailed geographic data for the census blocks, along with district data, population data, and partisan data.  These data tend  to come separately!  That is, we have Census data with census blocks and population.  One can also find partisan data in the form of election results, although those results are at the \emph{precinct} level (precincts are generally much larger than blocks, and blocks do not always nest nicely inside of precincts!).  And one can also find district data, though the State House, State Senate, and Congressional districts are all in different data files\footnote{We focus on those three maps for this project, although a state may have other districts.}.  These data must be coalesced into one file, which is as clean as possible.  Cleaning shapefiles for this purpose is an extremely nontrivial process; we describe that process in Section \ref{sec:construction}.  

\subsection{The Data We Used}\label{sec:data}

As stated in Section \ref{sec:oa_and_gerrychain}, in order to create the file to run an MCMC such as ReCom in gerrychain, we need many different types of data, cleaned, and coalesced together using the same geographic units.  These data used to be extremely difficult to obtain; we have anecdotally heard that, when requesting data from various states while putting together repositories that eventually became \cite{MGGGstates}, the MGGG was sent hand-drawn maps in some cases.

Amazingly, today one can access all of these kinds of necessary data from a single source: the Redistricting Data Hub (RDH)\cite{RDH}.  The RDH is a nonpartisan group that hosts a plethora of data for every state, and the list of available datasets is constantly growing.  There is no way that we could have created our datasets with our small group without the RDH's easily accessible datasets, and their partners such as VEST \cite{VEST} who allow the RDH to share their data.  

Additionally, there is no way that we could have cleaned and coalesced our datasets without the MGGG's maup Python library \cite{maup}.  This library was constructed specifically to deal with geospatial data used in redistricting.  The maup library includes many useful methods, but one of its most important methods for our purposes is {\tt smart\_repair()}.  When constructing the dual graph mentioned in Section \ref{sec:oa_and_gerrychain}, the geographic data must cover the state, and must have no gaps or overlaps between geographic units.  Many of the shapefiles that we needed to use for this project do have gaps and overlaps, and thus must be repaired.  The {\tt smart\_repair()} method does this in a ``smart'' way, so that the adjacencies between geographic units is appropriately assigned.  This repair method is incredibly complex, and we refer the reader to \cite{smart_repair} to learn more about it.  Needless to say, maup and {\tt smart\_repair()} were essential for this data cleaning project.

\section{The Repos}\label{sec:repositories}

The set of repositories that we constructed, one for each state\footnote{Not all states have been completed at the time of submission, but we plan to complete all 50 states.}, are available at \url{https://github.com/eveomett-states}.  For each state, we construct a json which can be used by the gerrychain Python library to run an MCMC process for further analysis.  We construct these jsons and make them publicly available so that researchers and students can study redistricting, and the public can use more robust MCMC tools such as gerrychain to study and analyze maps.  We were inspired by MGGG's repositories of maps drawn after the 2010 census \cite{MGGGstates}, and wanted to made a similar useful set of data from maps drawn after the 2020 census.  In Section \ref{sec:construction} we detail the steps taken to create these jsons, and in Section \ref{sec:available} we detail what is available in each state's repository.

\subsection{How We Built Them}\label{sec:construction}

In Section \ref{sec:oa_and_gerrychain}, we note that any districting map is constructed of partitions of the dual graph of Census blocks into connected components.  However, for most states, the number of Census blocks is so enormous that running a large ReCom chain on Census blocks is quite computationally intensive.  It is also generally not needed for a robust outlier analysis \cite{RecomMGGG}.  Thus, researchers and expert witnesses will commonly use precincts as the minimal geographic blocks that are combined to form districts.  For these reasons, all of our states' datasets we constructed are at the precinct level\footnote{We discuss some issues that arose with precincts in Section \ref{sec:caveats_conclusions}.}.

All of the data that we combined into a single json, we obtained from the Redistricting Data Hub \cite{RDH}.  These data include:
\begin{enumerate}
\item Census data (in 3 different datasets).  
\item Map data for State House, State Senate, and Congressional districts (or whatever subset of those the state in question has).
\item Election data that the RDH obtained from VEST \cite{VEST} from all of the elections 2016 or later that we could successfully clean.
\item County data from the Census (for the purpose of nesting precincts within counties, which we describe in more detail below).
\end{enumerate}
The election data is a proxy for partisan preference.  We only incorporate election results for state-wide races, so that the results are consistent by region, and straightforward to interpret.

We used MGGG's maup Python library \cite{maup} to both clean and coalesce these data.  The maup library has tools to:
\begin{enumerate}
\item Check if geographic data has gaps or overlaps.  
\item Clean geographic data that has gaps and/or overlaps between geographic units, in a way that is appropriate for the purposes of redistricting.  (Again, please see \cite{smart_repair} for details.)
\item Appropriately disaggregate election data from the precinct level to the block level.
\item Aggregate data from the block level to the precinct level.
\end{enumerate}
We used these capabilities, and constructed methods to do typical needed tasks, including:
\begin{enumerate}
\item Repair a dataset with gaps and/or overlaps
\item Initialize a single set of election data as the ``base'' data, to which other datasets are added, making sure to nest these election precincts within counties.
\item Add election data, appropriately prorating due to the different precincts used in different election years.
\item Add district data.
\item Rename columns so that they are consistently and understandably named.
\item Check boundary lengths of adjacent precincts, and mend small rook adjacencies to queen adjacencies.  
\end{enumerate}
We did not create a method to add population data, as this is simple enough that we felt it did not necessitate a new method.

We note that nesting precincts within counties (number 2 above) and mending small rook adjacencies (number 6 above) were recommended to us by the maup Python library developers as best practices.  Please see the maup documentation for details \cite{maup}.  Another best practice recommended by the maup developers was to use the Universal Transverse Mercator (UTM) coordinate reference system (CRS)\footnote{The crs tells how the 3-dimensional earth data is projected to get a 2-dimensional map.} for all data; it is needed to implement {\tt smart\_repair()}, and has higher precision than geographic crs's.  

For transparency purposes, we included the jupyter notebook used to create each state's json in that state's repository.  The notebook also contains code to construct the corresponding shapefile (which is much larger), which is also available in the repository if the zipped shapefile is not larger than 25 MB\footnote{The user can simply run the notebook to construct the shapefile if it is desired, but not included in that state's repository.}. The included notebook allows users to see both the steps taken in the data cleaning process, as well as the outputs we saw when running those data cleaning scripts.  Another reason we share this notebook is so that future users can easily incorporate additional election data as it becomes available.  In the methods that we constructed, we raised exceptions in a few cases, in order to halt the data cleaning process and alert the user that something may have gone wrong.  We did this when the built-in maup function {\tt maup.doctor()} reports that some gaps or overlaps remain after cleaning, and also when the population counts changed (indicating that some Census blocks were lost in the aggregation process).  

Not every part of the data cleaning process could be easily automated, as each state has different regional data reported along with the election data in the VEST election shapefiles.  Additionally, each different districting map had its own data attributes, which were not consistent map-to-map or state-to-state.  Thus, while we attempted to make the cleaning process as automated and easy as possible, future users will still need to be attentive to the unique datasets available to each state.

Finally, we note as a point of practical interest, the Census shapefiles were reliably extremely clean.  The shapefiles that required repairing in order to fix gaps and overlaps were by and large the election data files, and sometimes also the districting map files.

\subsection{What's Inside}\label{sec:available}

We are working on creating repos and jsons for every state; at the time of publication, we have completed 22 states \cite{eveomett-states}.   For each state's dataset, we included all VEST statewide election data from elections held from 2016 or later, so long as we could clean it sufficiently with the maup methods.  Only statewide election data is included, because other electoral races (such as US House elections) have different candidates on the ballot in different parts of the state, and even some uncontested races, which makes it a poor proxy for party preference.   We used consistent naming schemes for all of the election data, based on the naming schemes used by VEST.

Each state's json also includes a great deal of census data, including items like total population, population of varying demographic groups, voting age population, and voting age population by demographic group.  We include the same Census data as was available in the MGGG data from the post-2010-Census maps \cite{MGGGstates}, as well as the same naming schemes.

Finally, each state's json includes the State House, State Senate, and Congressional districting maps (or whatever subset of those that a state has).  To be more precise, some states have small enough population that they have a single Congressional district: the entire state.  We did not include the Congressional map for those states, since Congressional redistricting does not happen for such states.  All states except Nebraska have a lower house and upper house; since these are normally called the State Senate and State House, we used the same name ({\tt SEND}) for the State Senate (or upper house) and the same name ({\tt HDIST}) for the State House (or lower house).  For Nebraska's unicameral legislature, we simply used {\tt SEND} for its legislative districts.  Arizona, Idaho,  New Jersey, and Washington also just have a single state legislative map for state senate (their state house districts are multimember districts using the same map as the state senate).  {\tt CD} is used for a state's Congressional districting map; again following the conventions of \cite{MGGGstates}.

As mentioned in Section \ref{sec:construction}, we include in the repository the jupyter notebook used to create that state's data as a json\footnote{If the shapefile that was also constructed was not too large, it was also compressed and uploaded to that state's repository as a .zip file.}.  Thus, as new election data becomes available, the public should be able to relatively easily add that data to the combined dataset (json or shapefile) using the notebook and the {\tt add\_vest} method.  

\section{Complications, Caveats and Conclusions}\label{sec:caveats_conclusions}

We have cleaned and coalesced demographic, election, and district data into a single json file for 22 states so far \cite{eveomett-states}, and we plan to complete all 50 datasets by the year's end.   These jsons can be imported into the gerrychain Python library so that researchers, students, and the public can study maps from the post-2020-Census.  For transparency, as well as to allow others to reconstruct the jsons (and shapefiles) themselves or potentially add additional data, we include in each state's repository the jupyter notebook used to construct the data.

As mentioned in Section \ref{sec:construction}, while we constructed methods to make the data cleaning process as smooth as possible, there are enough issues with the data that it cannot be fully automatable. The more boring reasons are that each state's election file has the election data starting at a different column, and that the datasets have different names for the column giving the district name (including {\tt DISTRICT}, {\tt DISTRICTN}, {\tt DISTRICT\_I}, {\tt DISTRICTNO}, {\tt Name}, {\tt SENATE}, {\tt CONG\_DIST}, and {\tt ID}).  But there were more interesting issues that arose, such as Louisiana's state senate map.  Unlike the other shapefiles used for Louisiana, the senate districts had a hole across Lake Pontchartrain.  This corresponded to two geographic regions over the lake with no population and belonging to no state senate district, so we simply dropped those regions\footnote{It is worth noting that states sometimes do allow land regions within a district to be connected only across bodies of water, which this could potentially disconnect.  But we were fine with dropping those regions, since they didn't even exist in the senate shapefile.}

A potentially larger issue was the fact that maup assigns a census block's population to the precinct with which it has the largest overlap, and similarly assigns a precinct to the district with which it has the largest overlap.  This can introduce some error resulting in some district populations being inaccurate.  By and large, the assignments seemed reasonable, but for a couple of states this gives population deviations between districts that are larger than expected\footnote{We've noticed this in the State Senate districts in Hawaii, Nebraska, and Wyoming so far.}.  For such states, it may be worth constructing a single json file which is on the block level, rather than the precinct level.  Indeed, one future addition may be to add another json which is on the block level for researchers and students interested in that option.  Those files would be much larger in size (since blocks are smaller, and thus there are more of them), but it may be possible to run chains on the block level with enough computing power (or with future, faster code).  

While we have used the best possible tools to coalesce and clean these data, we do not give any kind of legal guarantee, and these data are not intended to be used for litigation purposes.  Our hope is that these data will allow researchers, students, and others easy access to current maps to further the study of redistricting and gerrymandering.

\section*{Acknowledgements}

We would like to give our utmost thanks to Jeanne Clelland, one of the maup Python library's developers.  We could not have completed this project without her help and advice.

This material is based upon work supported by the National Science Foundation under Grant No. DMS-1928930 and by the Alfred P. Sloan Foundation under grant G-2021-16778, while Ellen Veomett was in residence at the Simons Laufer Mathematical Sciences Institute (formerly MSRI) in Berkeley, California, during the Fall 2023 semester.

\bibliographystyle{plain}
\bibliography{DataCleaning}

\end{document}